\documentclass[prd,showpacs,showkeys,nofootinbib,preprint]{revtex4-1}

\usepackage{graphicx}
\usepackage{amsmath,amsfonts,amssymb}

\begin{document}

\title{Poincar\'e gauge gravity: An overview}

\author{Yuri N. Obukhov}
\email{obukhov@ibrae.ac.ru}
\affiliation{Russian Academy of Sciences, Nuclear Safety Institute,
B.Tulskaya 52, 115191 Moscow, Russia}

\begin{abstract}
We review the basics and the current status of the Poincar\'e gauge theory of gravity. The general dynamical scheme of Poincar\'e gauge gravity (PG) is formulated, and its physical consequences are outlined. In particular, we discuss exact solutions with and without torsion, highlight the cosmological aspects, and consider the probing of the spacetime geometry.
\end{abstract}


\maketitle

\section{Introduction: gauge symmetries, currents, and fields}

The gauge approach in field theory has a long history, going back to the early works of Weyl \cite{Weyl}, Cartan \cite{Cartan}, Fock \cite{Fock}, and later contributions by Utiyama \cite{Utiyama}, Sciama \cite{Sciama}, and Kibble \cite{Kibble}. The detailed review of the development of gauge gravity can be found in \cite{Hehl:1976,Trautman:1980,Ivanenko,Shapiro,Blagojevic,selected,Trautman:2006}, and especially complete and informative is the recent book \cite{Reader}. Here we give a brief overview of the subject, presenting the basic notions and mathematical structures and highlighting the physical consequences of the gauge theory of gravity based on the Poincar\'e symmetry group $G\!=\!T_4\!\rtimes\!SO(1,3)$. 

It is a nontrivial problem to extend the Yang-Mills \cite{Yang} approach of internal symmetry groups to those of spacetime symmetries. Without going into technical details, one can sketch the gauge-theoretic scheme as follows: The invariance of the action under an $N$-parameter group of field transformations yields, via the Noether theorem, $N$ conserved currents. When the parameters are allowed to be functions of spacetime coordinates, one needs to introduce $N$ gauge fields, which are coupled to the Noether currents, to preserve the invariance under the local (gauge) symmetry. In accordance with the general Yang-Mills-Utiyama-Kibble scheme, the 10-parameter Poincar\'e group gives rise to the 10-plet of the gauge potentials which are identified with the coframe $\vartheta^\alpha = e^\alpha_i dx^a$ (4 potentials corresponding to the translation subgroup $T_4$) and the local connection $\Gamma^{\alpha\beta} = -\,\Gamma^{\beta\alpha} = \Gamma_{i}{}^{\alpha\beta} dx^i$ (6 potentials for the Lorentz subgroup $SO(1,3)$). The ``translational'' and ``rotational'' field strengths then read 
\begin{eqnarray}
T^\alpha &=& D\vartheta^\alpha = d\vartheta^\alpha +\Gamma_\beta{}^\alpha\wedge
\vartheta^\beta,\label{Tor}\\ \label{Cur}
R^{\alpha\beta} &=& d\Gamma^{\alpha\beta} + \Gamma_\gamma{}^\beta\wedge\Gamma^{\alpha\gamma}.
\end{eqnarray}
They are interpreted as the torsion and the curvature 2-forms, thus naturally introducing the Riemann-Cartan geometry \cite{LdB} on the spacetime manifold.

These gravitational gauge fields are coupled to the Noether currents of the Poincar\'e group: the energy-momentum ${\mathfrak T}_\alpha$ and the spin ${\mathfrak S}_{\alpha\beta} = -\,{\mathfrak S}_{\beta\alpha}$. In a similar way, one can view Einstein's general relativity (GR) as the gauge theory based on the translation group $T_4$ with the coframe $\vartheta^\alpha$ as the gauge potential coupled to the energy-momentum ${\mathfrak T}_\alpha$ as the physical source of gravity \cite{Cho}. 

Our basic notation and conventions are as follows: Greek indices $\alpha, \beta,\dots{} = 0, \dots, 3$, denote the anholonomic components (for example, of a coframe $\vartheta^\alpha$), while the Latin indices $i,j,\dots{} =0,\dots, 3$, label the holonomic components ($dx^i$, e.g.). From the volume 4-form $\eta$, the $\eta$-basis is constructed with the help of the interior products as $\eta_{\alpha_1 \dots\alpha_p}:= e_{\alpha_p}\rfloor\dots e_{\alpha_1}\rfloor\eta$, $p=1,\dots,4$. These forms are related to the $\theta$-basis via the Hodge dual operator $^\ast$, for example, $\eta_\alpha = {}^\ast\vartheta_\alpha$ and $\eta_{\alpha\beta} = {}^\ast\left(\vartheta_\alpha\wedge\vartheta_\beta\right)$. The Minkowski metric $g_{\alpha\beta} = {\rm diag}(+1,-1,-1,-1)$ is used to lower and raise anholonomic indices: e.g., $e^\alpha = g^{\alpha\beta}e_\beta$. 
We {\it do not} use the natural units, and all the fundamental constants appear explicitly. In particular, the velocity of light $c$ factor is needed in many key formulas for dimensional reasons.

Only a limited number of references is given here; for a more complete bibliography on the Poincar\'e gauge gravity see the recent book \cite{Ponomarev}.

\subsection{Dynamical currents}

Let the matter field $\psi^A$ be a tensor-valued $p$-form. Its tensor structure is encoded in the multi-index \begin{footnotesize}$A$\end{footnotesize}, and dynamics is described by a general Lagrangian 4-form
\begin{eqnarray} 
L = L(\vartheta^\alpha\,, d\vartheta^\alpha\,,\Gamma^{\alpha\beta}\,, 
d\Gamma^{\alpha\beta}\,,\psi^A, d\psi^A) 
= L(\psi^A, D\psi^A, \vartheta^\alpha, T^\alpha, R^{\alpha\beta})\,.\label{L} 
\end{eqnarray}
The covariant derivative is defined by $D\psi^A = d\psi^A - {\frac 12}\Gamma^{\alpha\beta}\wedge(\rho^A{}_B)_{\alpha\beta}\psi^B$ with the Lorentz generators $(\rho^A{}_B)_{\alpha\beta} = -\,(\rho^A{}_B)_{\beta\alpha}$.

The {\it matter currents} are given by
\begin{eqnarray}
{\mathfrak T}_{\alpha} &:=& -\,{\frac {\delta L}{\delta\vartheta^{\alpha}}} =  
- \,{\frac {\partial L}{\partial\vartheta^{\alpha}}} 
- D\,{\frac {\partial L}{\partial T^{\alpha}}}\, ,\label{sigC0}\\
c{\mathfrak S}_{\alpha\beta} &:=& -\,2{\frac {\delta L}{\delta\Gamma^{\alpha\beta}}} =   
(\rho^A{}_B)_{\alpha\beta}\psi^B\wedge{\frac {\partial L} {\partial D\psi^A}}
- 2\vartheta_{[\alpha}\wedge {\frac {\partial L}{\partial T^{\beta]}}} 
-2D{\frac {\partial L}{\partial R^{\alpha\beta}}}\,.\label{spin0}
\end{eqnarray}

\subsection{Conservation laws}

\subsubsection{Diffeomorphism symmetry}

The invariance of $L$ under the local diffeomorphisms on the spacetime manifold yields the {\it first Noether identity}
\begin{eqnarray}
D{\mathfrak T}_\alpha &\equiv & (e_\alpha\rfloor T^\beta)\wedge{\mathfrak T}_\beta
+ {\frac 12}(e_\alpha\rfloor R^{\beta\gamma})\wedge c{\mathfrak S}_{\beta\gamma}
+\,W_{\alpha},\label{conmomC}
\end{eqnarray}
where the generalized force is $W_{\alpha} := -\,(e_\alpha\rfloor D\psi^A)\wedge{\frac{\delta L}{\delta\psi^A}} - (-1)^p(e_\alpha\rfloor\psi^A)\wedge D{\frac{\delta L}{\delta\psi^A}}$, with ${\frac {\delta L} {\delta\psi^A}} = {\frac {\partial  L}{\partial\psi^A}} - (-1)^{p}D\,{\frac {\partial L}{\partial (D\psi^A)}}$. As another consequence of the translational invariance one finds the explicit form of the {\it canonical energy--momentum current}:
\begin{eqnarray}
{\mathfrak T}_\alpha &=&  (e_\alpha\rfloor D\psi^A)\wedge {\frac{\partial L}{\partial D\psi^A}}
+ (e_\alpha\rfloor\psi^A)\wedge{\frac{\partial L}{\partial\psi^A}} - e_\alpha\rfloor L\nonumber\\
&& -\,D{\frac{\partial L}{\partial T^\alpha}} + (e_{\alpha}\rfloor T^\beta)\wedge
{\frac{\partial L}{\partial T^\beta}} + (e_{\alpha}\rfloor R^{\beta\gamma})\wedge 
{\frac{\partial L}{\partial R^{\beta\gamma}}}.\label{momC}
\end{eqnarray}

\subsubsection{Local Lorentz symmetry}

When the Lagrangian $L$ is invariant under the local Lorentz transformations
\begin{equation}\label{lor2}
\delta\vartheta^{\alpha} = \varepsilon_{\beta}{}^{\alpha}\,\vartheta^{\beta},\qquad 
\delta\Gamma^{\alpha\beta} = - D\varepsilon^{\alpha\beta},\qquad \delta
\psi^A = -\,{\frac 12}\varepsilon^{\alpha\beta}\,(\rho^A{}_B)_{\alpha\beta}\,\psi^B,
\end{equation}
with $\varepsilon^{\alpha\beta} = -\,\varepsilon^{\beta\alpha}$, we find the {\it second Noether identity}
\begin{equation}\label{Noe2}
cD{\mathfrak S}_{\alpha\beta} + \vartheta_\alpha\wedge
{\mathfrak T}_\beta - \vartheta_\beta\wedge{\mathfrak T}_\alpha \equiv W_{\alpha\beta}.
\end{equation}
The generalized torque is defined as $W_{\alpha\beta} := -\,(\rho^A{}_B)_{\alpha\beta}\psi^B\wedge {\frac{\delta L}{\delta\psi^A}}$.

\subsubsection{Gravitational Lagrangian and Noether identities}

The gravitational Lagrangian 4-form 
\begin{equation}
V = V(\vartheta^{\alpha}, T^{\alpha}, R^{\alpha\beta})\label{lagrV}
\end{equation}
is assumed to be an arbitrary function of the geometrical variables.

We introduce the {\it gauge field momenta} (``excitations'') 2-forms
\begin{equation} 
H_{\alpha} :=  c{\frac{\partial V}{\partial T^{\alpha}}}\,,\qquad  H_{\alpha\beta} := 
2{\frac{\partial V}{\partial R^{\alpha\beta}}}\, ,\label{HH}
\end{equation}
the {\it canonical} energy--momentum and spin $3$-forms for the Poincar\'e gauge fields
\begin{equation} 
E_{\alpha} := -\,c{\frac{\partial V}{\partial\vartheta^{\alpha}}},\qquad
E_{\alpha\beta} := -\,2{\frac{\partial V}{\partial\Gamma^{\alpha\beta}}} = 
- \,{\frac 2c}\vartheta_{[\alpha}\wedge H_{\beta]}\,, \label{EE}
\end{equation}
and find the variational derivatives with respect to the gravitational
field potentials
\begin{eqnarray}
{\mathcal E}_\alpha &:=& {\frac{\delta V}{\delta\vartheta^{\alpha}}} = 
{\frac 1c}\left(DH_{\alpha} - E_{\alpha}\right), \label{dVt}\\ 
{\mathcal C}_{\alpha\beta} &:=& {\frac{\delta V}{\delta\Gamma^{\alpha\beta}}} 
= {\frac 12}\left(DH_{\alpha\beta} - E_{\alpha\beta}\right).\label{dVG}
\end{eqnarray}
Diffeomorphism invariance yields the Noether identities
\begin{eqnarray}
E_{\alpha} &\equiv& -\,c\,e_{\alpha}\rfloor V + (e_{\alpha}\rfloor T^{\beta})\wedge H_{\beta} 
+ {\frac c2}(e_{\alpha}\rfloor R^{\beta\gamma})\wedge H_{\beta\gamma},\label{Ea}\\
D\,{\mathcal E}_\alpha &\equiv& (e_{\alpha}\rfloor T^{\beta})\wedge{\mathcal E}_\beta 
+ (e_{\alpha}\rfloor R^{\beta\gamma})\wedge\,{\mathcal C}_{\beta\gamma},\label{1st}
\end{eqnarray}
whereas the local Lorentz invariance results in the Noether identity
\begin{equation}
2D{\mathcal C}_{\alpha\beta} + \vartheta_\alpha\wedge{\mathcal E}_\beta 
- \vartheta_\beta\wedge{\mathcal E}_\alpha \equiv 0\,.\label{2nd}
\end{equation}

\section{Mathematical interlude: irreducible decompositions}

\subsection{Torsion decomposition}

The torsion 2-form can be decomposed into the three irreducible pieces, $T^{\alpha}={}^{(1)}T^{\alpha} + {}^{(2)}T^{\alpha} + {}^{(3)}T^{\alpha}$, where
\begin{eqnarray}
{}^{(2)}T^{\alpha} &=& {\frac 13}\vartheta^{\alpha}\wedge T,\qquad 
{}^{(3)}T^{\alpha} = {\frac 13}e^\alpha\rfloor{}^\ast \overline{T},\label{iT23}\\
{}^{(1)}T^{\alpha} &=& T^{\alpha}-{}^{(2)}T^{\alpha} - {}^{(3)}T^{\alpha}.\label{iT1}
\end{eqnarray}
Here the 1-forms of the torsion trace and axial trace are introduced:
\begin{equation}
T := e_\nu\rfloor T^\nu,\qquad \overline{T} := {}^*(T^{\nu}\wedge\vartheta_{\nu}).\label{traces1}
\end{equation}

For the irreducible pieces of the dual torsion ${}^*T^{\alpha} = {}^{(1)}({}^*T^{\alpha}) + {}^{(2)}({}^*T^{\alpha}) + {}^{(3)}({}^*T^{\alpha})$, we have the properties 
\begin{equation}
{}^{(1)}({}^*T^\alpha)={}^*({}^{(1)}T^\alpha),\quad
{}^{(2)}({}^*T^\alpha)={}^*({}^{(3)}T^\alpha),\quad
{}^{(3)}({}^*T^\alpha)={}^*({}^{(2)}T^\alpha).\label{dTdual}
\end{equation}

\subsection{Curvature decomposition}

The Riemann-Cartan curvature 2-form is decomposed $R^{\alpha\beta} = \sum_{I=1}^6\,{}^{(I)}\!R^{\alpha\beta}$ into the 6 irreducible parts 
\begin{eqnarray}
&{}^{(2)}\!R^{\alpha\beta} = -\,{}^*(\vartheta^{[\alpha}\wedge\overline{\Psi}{}^{\beta]}),\qquad
{}^{(4)}\!R^{\alpha\beta} = -\,\vartheta^{[\alpha}\wedge\Psi^{\beta]},& \label{curv24}\\
&{}^{(3)}\!R^{\alpha\beta} = -\,{\frac 1{12}}\,\overline{X}\,{}^*\!(\vartheta^\alpha\wedge
\vartheta^\beta),\qquad {}^{(6)}\!R^{\alpha\beta}  = -\,{\frac 1{12}}\,X\,\vartheta^\alpha\wedge
\vartheta^\beta,& \label{curv36}\\
&{}^{(5)}\!R^{\alpha\beta} = -\,{\frac 12}\vartheta^{[\alpha}\wedge e^{\beta]}
\rfloor(\vartheta^\gamma\wedge X_\gamma),& \label{curv5}\\
&{}^{(1)}\!R^{\alpha\beta} = R^{\alpha\beta} - \sum_{I=1}^6\,{}^{(I)}\!R^{\alpha\beta},& \label{curv1}
\end{eqnarray}
where 
\begin{equation}
X^\alpha := e_\beta\rfloor R^{\alpha\beta},\quad X := e_\alpha\rfloor X^\alpha,
\quad \overline{X}^\alpha := {}^*(R^{\beta\alpha}\wedge\vartheta_\beta),\quad 
\overline{X} := e_\alpha\rfloor \overline{X}^\alpha,\label{WX}
\end{equation}
and 
\begin{eqnarray}
&\Psi_\alpha := X_\alpha - {\frac 14}\,\vartheta_\alpha\,X - {\frac 12}
\,e_\alpha\rfloor (\vartheta^\beta\wedge X_\beta),&\label{Psia}\\
&\overline{\Psi}_\alpha := \overline{X}_\alpha - {\frac 14}\,\vartheta_\alpha
\,\overline{X} - {\frac 12}\,e_\alpha\rfloor (\vartheta^\beta\wedge \overline{X}_\beta).&\label{Phia}
\end{eqnarray}
The 1-forms $X^\alpha$ and $\overline{X}^\alpha$ are not completely independent: $\vartheta_\alpha\wedge X^\alpha = {}^*(\vartheta_\alpha\wedge \overline{X}^\alpha)$.

The curvature tensor $R_{\mu\nu}{}^{\alpha\beta}$ is constructed from the components of the 2-form $R^{\alpha\beta} = {\frac 12}R_{\mu\nu}{}^{\alpha\beta}\,\vartheta^\mu\wedge\vartheta^\nu$. The Ricci tensor is defined as ${\rm Ric}_{\alpha}{}^{\beta} := R_{\gamma\alpha}{}^{\beta\gamma}$. The curvature scalar $R = {\rm Ric}_{\alpha}{}^{\alpha}$ determines the 6-th irreducible part since $X \equiv R$. The first irreducible part (\ref{curv1}) introduces the generalized Weyl tensor $C_{\mu\nu}{}^{\alpha\beta}$ via the expansion of the 2-form ${}^{(1)}R^{\alpha\beta} = {\frac 12}C_{\mu\nu}{}^{\alpha\beta}\,\vartheta^\mu\wedge\vartheta^\nu$. From (\ref{Phia}) we learn that the 4-th part of the curvature is given by the symmetric traceless Ricci tensor,   
\begin{equation}
\Psi_\alpha = \Bigl({\rm Ric}_{(\alpha\beta)} - {\frac 14}\,R\,g_{\alpha\beta}
\Bigr)\vartheta^\beta.\label{RicP}
\end{equation}
Accordingly, the 1-st, 4-th and 6-th curvature parts generalize the well-known irreducible decomposition of the Riemannian curvature tensor. The 2-nd, 3-rd and 5-th curvature parts are purely non-Riemannian.

\section{Matter sources in Poincar\'e gravity}

Matter with spin is the source of gravity in PG theory. Here we specify two explicit 
examples of macro- and microscopic origin.

\subsection{Macroscopic matter: spinning fluid}

Weyssenhoff's fluid \cite{Weyss} represents a special case of a medium with microstructure. To describe its dynamics, one rigidly attaches a triad $b^\alpha_A$, $A = 1,2,3$, to matter elements. It is orthogonal to fluid's flow that is represented by the flow 3-form $u$.

The physical properties of the fluid are described by the particle density $\rho$, the entropy $s$, the specific (per matter element) spin density $\mu^{AB} = - \mu^{BA}$, and the internal energy density $\varepsilon = \varepsilon(\rho, s, \mu^{AB})$. The Gibbs law of thermodynamics is corrected by the contribution of the spin energy:
\begin{equation}
Tds = d \left({\frac \varepsilon \rho}\right) + p\,d\left({\frac 1 \rho}\right) 
- {\frac 12}\omega_{AB}d\mu^{AB}.\label{Tds}
\end{equation}
Here $T$ is the temperature, $p$ is the pressure, and $\omega_{AB}$ is the thermodynamical variable conjugated to the specific spin density $\mu^{AB}$.

We assume that the fluid moves such that the particle number is not changed and the entropy and identity of fluid elements is preserved along the lines of flow:
\begin{eqnarray}
d(\rho u) = 0,\qquad u\wedge ds = 0,\qquad u\wedge dX = 0,\label{udx}
\end{eqnarray}
where $X$ is Lin's identity variable. 

The Lagrangian 4-form of the spinning fluid reads \cite{OK}
\begin{equation}
L = -\,\varepsilon\eta + {\frac 12}\rho\mu^{AB}g_{\alpha\beta}b^\alpha_A u\wedge Db^\beta_B +
L_{\rm con},\label{LW}
\end{equation}
where the constraints are imposed on the flow by means of the Lagrange multipliers:
\begin{eqnarray}
L_{\rm con} &=& \lambda_0({}^\ast u \wedge u - c^2\eta) + \lambda^Ab^\alpha_A \vartheta_\alpha\wedge u 
+ \lambda^{AB}(g_{\alpha\beta}b^\alpha_A b^\beta_B + \delta_{AB})\eta \nonumber\\
&& -\,\rho u\wedge d\lambda_1 + \lambda_2 u\wedge ds + \lambda_3 u\wedge dX.\label{Lc}
\end{eqnarray}
Variation with respect to $\lambda_0, \lambda^A, \lambda^{AB}, \lambda_1, \lambda_2, \lambda_3$ yields (\ref{udx}) and the orthogonality and normalization constraints for the flow 3-form $u$ and the material triad $b^\alpha_A$.

The canonical energy-momentum and spin currents (\ref{sigC0}) and (\ref{spin0}) are found as
\begin{eqnarray}\label{TW}
{\mathfrak T}_{\alpha} &=& u{\mathcal P}_\alpha - p\Big(\eta_\alpha - {\frac 1{c^2}}u_\alpha u\Big),\\
c{\mathfrak S}_{\alpha\beta} &=& u\,{\mathcal S}_{\alpha\beta}.\label{SW} 
\end{eqnarray}
Here  $u_\alpha = e_\alpha\rfloor {}^\ast u$, and the 4-momentum density and the covariant spin density of the medium are introduced by
\begin{eqnarray}\label{PaSab}
{\mathcal P}^\alpha = {\frac 1{c^2}}\Big[\varepsilon u^\alpha - u_\beta{}^\ast\!
D(u\,{\mathcal S}^{\alpha\beta})\Big],\qquad {\mathcal S}^{\alpha\beta} = -\,\rho\mu^{AB}b^\alpha_Ab^\beta_B.
\end{eqnarray}
The covariant spin density satisfies the Frenkel supplementary condition $u^\beta{\mathcal S}_{\alpha\beta} = 0$ (by construction) and its dynamics is governed by the equation of motion
\begin{equation}
D(u\,{\mathcal S}_{\alpha\beta}) - {\frac 1{c^2}}u_\beta u^\gamma D(u\,{\mathcal S}_{\alpha\gamma})
- {\frac 1{c^2}}u_\alpha u^\gamma D(u\,{\mathcal S}_{\gamma\beta}) = 0.\label{spineq}
\end{equation}
Note that the quadratic spin scalar invariant is conserved,
\begin{equation}\label{dS0}
d({\mathcal S}u) = 0,\qquad {\mathcal S}^2 = {\frac 12}\,{\mathcal S}_{\mu\nu}{\mathcal S}^{\mu\nu},
\end{equation}
which is an immediate consequence of (\ref{spineq}).

\subsection{Microscopic matter: Dirac spinor field}

The Dirac spin ${\frac 12}$ field is most conveniently discussed in the formalism of Clifford algebra-valued exterior forms, when the basic objects are the matrix-valued one- or three-forms $\gamma = \gamma_\alpha\,\vartheta^\alpha$ and ${}^\ast\gamma=\gamma^\alpha\,\eta_\alpha$. Unlike the usual 1-forms, such objects do not anticommute; in particular, ${\frac i{4!}}\gamma\wedge\gamma\wedge\gamma\wedge\gamma = \gamma_5\eta$. 

The Lagrangian 4-form of a Dirac field $\Psi$ is given by 
\begin{equation}
L_{\rm D} = -\,{\frac{i}{2}}\hbar c\left\{\overline{\Psi}\,{}^\ast\gamma\wedge D\Psi
+\overline{D\Psi}\wedge{}^\ast\gamma\,\Psi\right\} - {}^\ast mc^2\,
\overline{\Psi}\Psi\,.\label{lagr}
\end{equation}
The Dirac-conjugate spinors are denoted by $\overline{\Psi}$. Geometrically, Dirac fields are local sections of the spinor $SO(1,3)$-bundle associated with the principal bundle of orthonormal frames, so that the spinor covariant derivative reads $D\Psi = d\Psi + \frac{i}{4}\Gamma^{\alpha\beta}\wedge\sigma_{\alpha\beta}\,\Psi$, where the Lorentz algebra generators are $\sigma_{\alpha\beta} = i\gamma_{[\alpha}\gamma_{\beta]}$.

The Dirac wave equation is derived from the variation of the action with respect to the 
spinor field:
\begin{eqnarray}
i\hbar{}^\ast\gamma\wedge \left(D\,\Psi - \hbox{$\scriptstyle\frac{1}{2}$}
T\,\Psi\right) + {}^\ast mc\,\Psi = 0.\label{dirRCa}
\end{eqnarray}
For the canonical energy-momentum and spin currents (\ref{momC}) and (\ref{spin0}) we find
\begin{eqnarray}\label{Sa}
{\mathfrak T}_\alpha &=& {\frac {i\hbar c}{2}}\left(\overline{\Psi}\,{}^\ast\!\gamma 
D_\alpha\Psi - D_\alpha\overline{\Psi}\,{}^\ast\!\gamma\Psi\right),\\
{\mathfrak S}_{\alpha\beta} &=& {\frac {\hbar}{4}}\,\overline{\Psi}\left(
\sigma_{\alpha\beta}\,{}^\ast\gamma + {}^\ast\gamma\,\sigma_{\alpha\beta}\right)\Psi
= {\frac {\hbar}{2}}\,\vartheta_\alpha\wedge\vartheta_\beta
\wedge\overline{\Psi}\gamma\gamma_5\Psi.\label{Tau}
\end{eqnarray}
Hereafter $D_\alpha:=e_\alpha\rfloor D$. A characteristic feature of the Dirac fermion is the completely antisymmetric spin (\ref{Tau}). This means that only an axial part of the spacetime torsion interacts with the Dirac spinor field.

\section{Poincar\'e gravity field equations}

The field equations for the system of interacting matter and gravitational fields are derived from the total Lagrangian
\begin{equation}
V(\vartheta^{\alpha}, T^{\alpha}, R^{\alpha\beta}) + {\frac 1c}
L(\psi^A, D\psi^A, \vartheta^{\alpha}, T^{\alpha}, R^{\alpha\beta}).\label{Ltot}
\end{equation}
Independent variation with respect to $\psi^A$, $\vartheta^\alpha$, and $\Gamma^{\alpha\beta}$ yields
\begin{eqnarray}
{\frac {\partial  L}{\partial\psi^A}} - (-1)^{p}D\,
{\frac {\partial L}{\partial (D\psi^A)}} = 0\,,&&\label{Pmat}\\ 
DH_{\alpha} - E_{\alpha} = {\mathfrak T}_{\alpha}\,,&&\label{Peq1}\\
DH_{\alpha\beta} - E_{\alpha\beta} = {\mathfrak S}_{\alpha\beta}\,.&&\label{Peq2}
\end{eqnarray} 
The factor $1/c$ in (\ref{Ltot}) is explained by the dimensional reasons.

By expanding the currents with respect to the $\eta$-basis, we find the energy-momentum tensor and the spin density tensor: ${\mathfrak T}_\alpha = {\mathfrak T}_\alpha{}^\mu\eta_\mu$, and ${\mathfrak S}_{\alpha\beta} = {\mathfrak S}_{\alpha\beta}{}^\mu\eta_\mu$.

\subsection{Einstein-Cartan model}

The Einstein-Cartan theory \cite{Trautman:2006} is based on the Hilbert-Einstein Lagrangian
\begin{equation}
V_{\rm HE} = {\frac {1}{2\kappa c}}\eta_{\alpha\beta}\wedge R^{\alpha\beta}.\label{LHE}
\end{equation}
Here $\kappa = {\frac {8\pi G}{c^4}}$ is Einstein's gravitational constant with the dimension of $[\kappa] = \,$N$^{-1}=\,$s$^2$\,kg$^{-1}$\,m$^{-1}$. Newton's gravitational constant is $G = 6.67\times 10^{-11}$ m$^3$\,kg$^{-1}$\,s$^{-2}$. The velocity of light $c = 2.9\times 10^8$ m/s. 

For the Lagrangian (\ref{LHE}) we find from (\ref{HH}), (\ref{EE}) and (\ref{Ea}):
\begin{equation}
H_\alpha = 0,\quad H_{\alpha\beta} = {\frac {1}{\kappa c}}\eta_{\alpha\beta},\quad
E_\alpha = -\,{\frac {1}{2\kappa}}\eta_{\alpha\beta\gamma}\wedge R^{\beta\gamma},
\quad E_{\alpha\beta} = 0.\label{HHEE}
\end{equation}
As a result, the Einstein-Cartan field equations read
\begin{equation}
 {\frac {1}{2}}\eta_{\alpha\beta\gamma}\wedge R^{\beta\gamma} = \kappa\,{\mathfrak T}_\alpha,\qquad
\eta_{\alpha\beta\gamma}\wedge T^{\gamma} = \kappa c\,{\mathfrak S}_{\alpha\beta}.\label{ECeq}
\end{equation}

Substituting $R^{\alpha\beta} = {\frac 12}R_{\mu\nu}{}^{\alpha\beta}\,\vartheta^\mu\wedge\vartheta^\nu$ and $T^{\alpha} = {\frac 12}T_{\mu\nu}{}^\alpha\,\vartheta^\mu\wedge\vartheta^\nu$ into (\ref{ECeq}), we find the Einstein-Cartan field equations in components
\begin{eqnarray}\label{EC1}
{\rm Ric}_\alpha{}^\beta - {\frac 12}\delta_\alpha^\beta\,R &=& \kappa\,{\mathfrak T}_\alpha{}^\beta,\\
T_{\alpha\beta}{}^\gamma - \delta_\alpha^\gamma T_{\mu\beta}{}^\mu + \delta_\beta^\gamma T_{\mu\alpha}{}^\mu
&=& \kappa c\,{\mathfrak S}_{\alpha\beta}{}^\gamma.\label{EC2}
\end{eqnarray}

\subsection{Quadratic Poincar\'e gravity models}

The general quadratic model is described by the Lagrangian 4-form that contains all possible quadratic invariants of the torsion and the curvature:
\begin{eqnarray}
V &=& {\frac {1}{2\kappa c}}\Big\{\Big(a_0\eta_{\alpha\beta} + \overline{a}_0
\vartheta_\alpha\wedge\vartheta_\beta\Big)\wedge R^{\alpha\beta} - 2\lambda_0\eta \nonumber\\
&& -\,T^\alpha\wedge\sum_{I=1}^3
\left[a_I\,{}^*({}^{(I)}T_\alpha) + \overline{a}_I\,{}^{(I)}T_\alpha\right]\Big\}\nonumber\\
&& - \,{\frac 1{2\rho}}R^{\alpha\beta}\wedge\sum_{I=1}^6 \left[b_I\,{}^*({}^{(I)}\!R_{\alpha\beta}) 
+ \overline{b}_I\,{}^{(I)}\!R_{\alpha\beta}\right].\label{LRT}
\end{eqnarray}
The Lagrangian has a clear structure: the first line is {\it linear} in the curvature, the second line collects {\it torsion quadratic} terms, whereas the third line contains the {\it curvature quadratic} invariants. Furthermore, each line is composed of the parity even pieces (first terms on each line), and the parity odd parts (last terms on each line). The dimensionless constant $\overline{a}_0 = {\frac 1\xi}$ is inverse to the so-called Barbero-Immirzi parameter $\xi$, and the linear part of the Lagrangian -- the first line in (\ref{LRT}) -- describes what is known in the literature as the Einstein-Cartan-Holst model. A special case $a_0 = 0$ and $\overline{a}_0 = 0$ describes the purely quadratic model without the Hilbert-Einstein linear term in the Lagrangian. In the Einstein-Cartan model, one puts $a_0 = 1$ and $\overline{a}_0 = 0$. 

Besides that, the general PG model contains a set of the coupling constants which determine the structure of quadratic part of the Lagrangian: $\rho$, $a_1, a_2, a_3$ and $\overline{a}_1, \overline{a}_2, \overline{a}_3$, $b_1, \cdots, b_6$ and $\overline{b}_1, \cdots, \overline{b}_6$. The overbar denotes the constants responsible for the parity odd interaction. We have the dimension $[{\frac 1\rho}] = [\hbar]$, whereas $a_I$, $\overline{a}_I$, $b_I$ and $\overline{b}_I$ are dimensionless. Moreover, not all of these constants are independent: we take $\overline{a}_2 = \overline{a}_3$, $\overline{b}_2 = \overline{b}_4$ and $\overline{b}_3 = \overline{b}_6$ because some of terms in (\ref{LRT}) are the same,
\begin{eqnarray}\label{T23}
T^\alpha\wedge{}^{(2)}T_\alpha = T^\alpha\wedge{}^{(3)}T_\alpha = {}^{(2)}T^\alpha\wedge{}^{(3)}T_\alpha,\\
R^{\alpha\beta}\wedge{}^{(2)}\!R_{\alpha\beta} = R^{\alpha\beta}\wedge{}^{(4)}\!R_{\alpha\beta} 
= {}^{(2)}\!R^{\alpha\beta}\wedge{}^{(4)}\!R_{\alpha\beta},\label{R24} \\ 
R^{\alpha\beta}\wedge{}^{(3)}\!R_{\alpha\beta} = R^{\alpha\beta}\wedge{}^{(6)}\!R_{\alpha\beta} 
= {}^{(3)}\!R^{\alpha\beta}\wedge{}^{(6)}R_{\alpha\beta}.\label{R36}
\end{eqnarray}

For the Lagrangian (\ref{LRT}) from (\ref{HH})-(\ref{EE}) we derive the gravitational field momenta 
\begin{eqnarray}
H_\alpha = -\,{\frac 1{\kappa}}\,h_\alpha\,,\qquad H_{\alpha\beta} = {\frac {1}{\kappa c}}
\left(a_0\,\eta_{\alpha\beta} + \overline{a}_0\vartheta_\alpha\wedge\vartheta_\beta\right)
- {\frac 2\rho}\,h_{\alpha\beta},\label{HabRT}
\end{eqnarray}
and the canonical energy-momentum and spin currents of the gravitational field
\begin{eqnarray}
E_\alpha &=& -\,{\frac {1}{\kappa}}\Big({\frac {a_0}2}\,\eta_{\alpha\beta\gamma}\wedge R^{\beta\gamma}
+ \overline{a}_0\,R_{\alpha\beta}\wedge\vartheta^\beta 
- \lambda_0\eta_\alpha + q^{(T)}_\alpha\Big) - {\frac c\rho}\,q^{(R)}_\alpha,\label{EaRT}\\
E_{\alpha\beta} &=& {\frac 1c}\left(H_\alpha\wedge\vartheta_\beta 
- H_\beta\wedge\vartheta_\alpha\right).\label{EabRT}
\end{eqnarray}
For convenience, we introduced here the 2-forms which are linear functions of the torsion and the curvature, respectively, by
\begin{eqnarray}
h_\alpha = \sum_{I=1}^3\left[a_I\,{}^*({}^{(I)}T_\alpha) 
+ \overline{a}_I\,{}^{(I)}T_\alpha\right],\quad 
h_{\alpha\beta} = \sum_{I=1}^6\left[b_I\,{}^*({}^{(I)}\!R_{\alpha\beta}) 
+ \overline{b}_I\,{}^{(I)}\!R_{\alpha\beta}\right],\label{hR}
\end{eqnarray}
and the 3-forms quadratic in the torsion and in the curvature, respectively:
\begin{eqnarray}
q^{(T)}_\alpha &=& {\frac 12}\left[(e_\alpha\rfloor T^\beta)\wedge h_\beta - T^\beta\wedge 
e_\alpha\rfloor h_\beta\right],\label{qa}\\
q^{(R)}_\alpha &=& {\frac 12}\left[(e_\alpha\rfloor R^{\beta\gamma})\wedge h_{\beta\gamma} 
- R^{\beta\gamma}\wedge e_\alpha\rfloor h_{\beta\gamma}\right].\label{qaR}
\end{eqnarray}
By construction, the first 2-form in (\ref{hR}) has the dimension of a length, $[h_\alpha] = [\ell]$, whereas the second one is obviously dimensionless, $[h_{\alpha\beta}] = 1$. Similarly, we find for (\ref{qa}) the dimension of length $[q^{(T)}_\alpha] = [\ell]$, and the dimension of the inverse length, $[q^{(R)}_\alpha] = [1/\ell]$ for (\ref{qaR}). 

The resulting Poincar\'e gravity field equations (\ref{dVt}) and (\ref{dVG}) then read:
\begin{eqnarray}
{\frac {a_0}2}\eta_{\alpha\beta\gamma}\wedge R^{\beta\gamma} + \overline{a}_0R_{\alpha\beta}
\wedge\vartheta^\beta - \lambda_0\eta_\alpha && \nonumber\\ 
+ \,q^{(T)}_\alpha + \ell_\rho^2\,q^{(R)}_\alpha - Dh_\alpha &=& \kappa\,{\mathfrak T}_\alpha,\label{ERT1}\\
a_0\,\eta_{\alpha\beta\gamma}\wedge T^{\gamma} + \overline{a}_0\left(T_\alpha\wedge\vartheta_\beta
- T_\beta\wedge\vartheta_\alpha \right) && \nonumber\\ 
+ \,h_\alpha\wedge\vartheta_\beta - h_\beta\wedge\vartheta_\alpha 
- 2\ell_\rho^2\,Dh_{\alpha\beta} &=& \kappa c\,{\mathfrak S}_{\alpha\beta}.
\label{ERT2}
\end{eqnarray}
The contribution of the curvature square terms in the Lagrangian (\ref{LRT}) to the gravitational field dynamics in the equations (\ref{ERT1}) and (\ref{ERT2}) is characterized by the new coupling parameter with the dimension of the area (recall that $[{\frac 1\rho}] = [\hbar]$):
\begin{equation}
\ell_\rho^2 = {\frac {\kappa c}{\rho}}.\label{lr}
\end{equation}
The parity-odd sector in PG gravity has been recently analysed in \cite{Diakonov,Baekler1,Baekler2,Chen,Ho2,Ho3}, with a particular attention to the cosmological issues.

\section{Classical solutions of PG theory}

Although numerous classical exact and approximate solutions are known in Poincar\'e gravity theory (see \cite{Mccrea2,Hehl3,OPZ} for a review), their existence and structure depend essentially on the choice of the Lagrangian. Instead of analyzing special models on the case by case basis, we discuss here the results which are established for the general quadratic PG model (\ref{LRT}).

\subsection{Generalized Birkhoff theorem}

Spherically symmetric solutions are of particular interest in field-theoretic models. In Einstein's GR the Schwarzschild solution is unique, which is a remarkable theoretical result known as the Birkhoff theorem. The validity of this theorem is very important since the fundamental gravitational experiments in our Solar system are perfectly consistent with the Schwarzschild geometry. 

In contrast to GR, a spherically symmetric solution is not unique in a general quadratic PG gravity theory. However, certain classes of models do admit the {\it generalized Birkhoff theorem} which can be formulated as follows: the Schwarzschild spacetime without torsion is unique vacuum spherically symmetric solution of Poincar\'e field equations. 

This theorem is available in two versions. In the weak version, the spherical symmetry is understood as the form-invariance of the geometrical variables under the $SO(3)$ group of rotations, whereas in the strong $O(3)$ version one assumes the invariance under the rotations {\it and} spatial reflections. 

The analysis of the validity of the generalized Birkhoff theorem in PG is based on the appropriate ansatz for the metric and the torsion. In the local coordinates $(t, r, \theta, \varphi)$, the most general spherically symmetric spacetime interval reads
\begin{equation}
ds^2 = A^2dt^2 - B^2dr^2 - C^2(d\theta^2 + \sin^2\theta d\varphi^2),\label{Sds}
\end{equation}
so that the coframe can be chosen in the form
\begin{equation}\label{Scof}
\vartheta^{\widehat 0} = Adt,\qquad \vartheta^{\widehat 1} = Bdr,\qquad
\vartheta^{\widehat 2} = Cd\theta,\qquad \vartheta^{\widehat 3} = C\sin\theta d\varphi.
\end{equation}
The three functions $A = A(t,r)$,  $B = B(t,r)$,  $C = C(t,r)$, may depend arbitrarily
on the time $t$ and the radial coordinate $r$.

Let us divide the anholonomic indices, $\alpha,\beta,\dots$, into the two groups: $A,B,\dots = 0,1$ and $a,b,\dots = 2,3$. Then the spherically symmetric torsion ansatz for its three irreducible parts can be written as follows:
\begin{eqnarray}
{}^{(1)}\!T^A = 2\vartheta^A\wedge V + 2e^A\rfloor{}^\ast\overline{V},&\qquad&
{}^{(1)}\!T^a = -\,\vartheta^a\wedge V - e^a\rfloor{}^\ast\overline{V},\label{ST1}\\
{}^{(2)}\!T^\alpha = {\frac 13}\vartheta^\alpha\wedge T,&\qquad&
{}^{(3)}\!T^\alpha = {\frac 13}e^\alpha\rfloor {}^\ast\overline{T}.\label{ST23}
\end{eqnarray}
Here the torsion trace 1-form $T$ and the axial torsion 1-form $\overline{T}$ are
\begin{equation}
T = u_A \vartheta^A,\qquad \overline{T} = \overline{u}_A\vartheta^A,\label{STT}
\end{equation}
whereas the traceless 1st irreducible torsion is constructed from the 1-forms
\begin{equation}
V = v_A \vartheta^A,\qquad \overline{V} = \overline{v}_A\vartheta^A.\label{SVV}
\end{equation}
All together, the general spherically symmetric ansatz for the torsion thus includes eight
variables -- the components of the 1-forms $T, \overline{T}, V, \overline{V}$:
\begin{equation}
u_A(t,r),\qquad \overline{u}_A(t,r),\qquad v_A(t,r),\qquad \overline{v}_A(t,r),\qquad
A = 0,1.\label{uv}
\end{equation} 
As usual, the overline denotes the parity-odd objects. All 8 torsion functions (\ref{uv}) are allowed in the discussion of the weak $SO(3)$ version of the generalized Birkhoff theorem, however, in the strong $O(3)$ version the parity-odd variables $\overline{u}_A =\overline{v}_A = 0$ (hence $\overline{T} = \overline{V} = 0$), reducing the number of nontrivial torsion components to 4.
\begin{center}
\begin{figure}
\includegraphics[width=13cm]{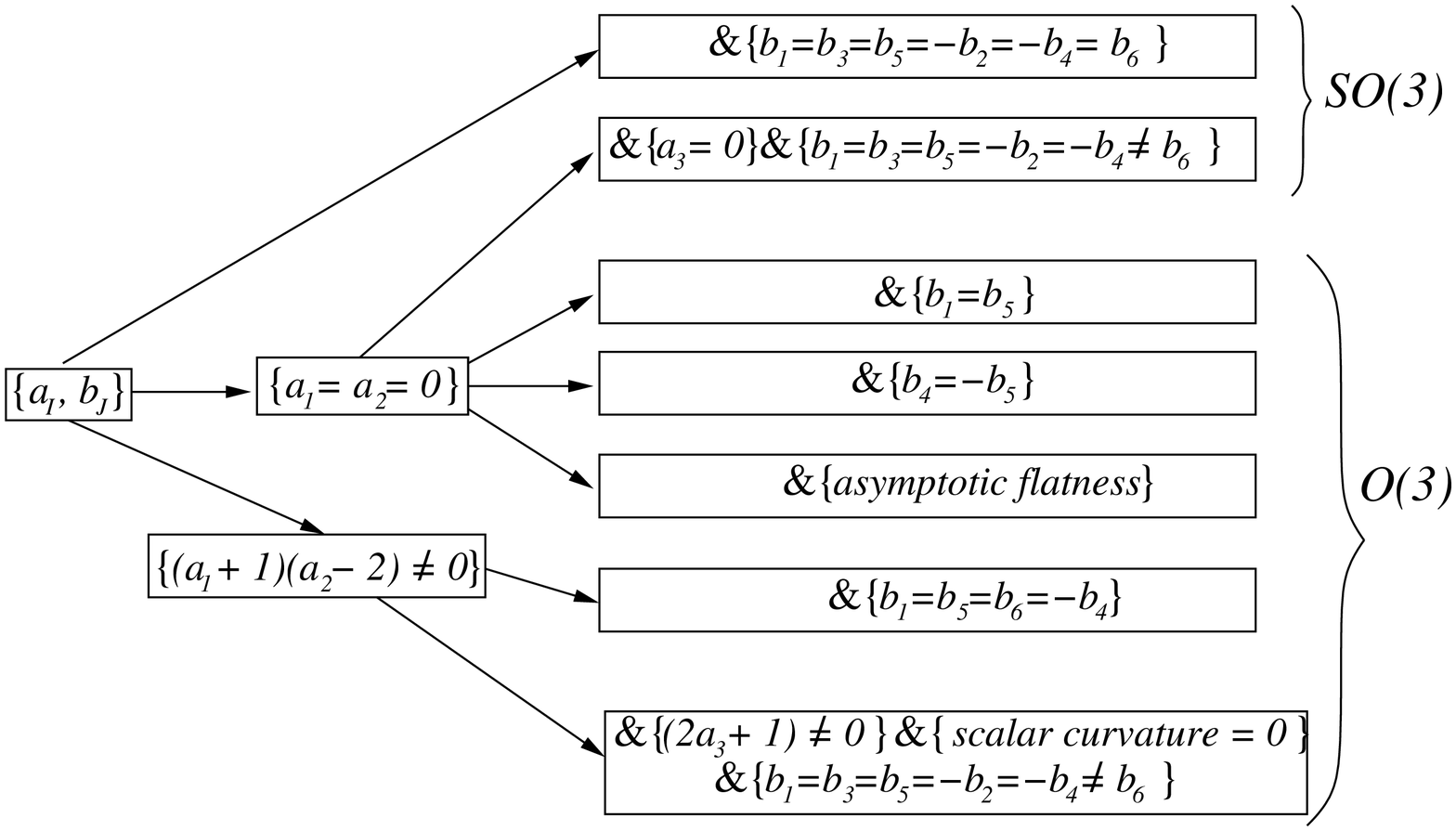}
\caption{The sufficient conditions for the generalized Birkhoff theorem -- in the weak $SO(3)$ version or in the strong $O(3)$ version. The simultaneously imposed conditions are linked by the symbol ``\&'' and by the arrows.}
\end{figure}
\end{center}

To prove the generalized Birkhoff theorem, one needs to plug the spherically symmetric ansatz (\ref{Sds})-(\ref{SVV}) into the field equations (\ref{ERT1})-(\ref{ERT2}) and to find the conditions under which these field equations yield the vanishing torsion and the reduction of the metric to the Schwarzschild form. Some of these conditions may impose constraints on the coupling constants, other conditions may impose constraints on the geometric structure. Among the latter assumptions are: (i) the asymptotic flatness condition which requires that the metric (\ref{Sds}) approaches the Minkowski line element, i.e. $A \longrightarrow 1$, $B \longrightarrow 1$, $C \longrightarrow r$ in the limit of $r\longrightarrow\infty$, or (ii) the vanishing scalar curvature $X = R = e_\alpha\rfloor e_\beta\rfloor R^{\alpha\beta} = 0$ condition. 

In the literature \cite{Rama,Rauch,OPZ}, only the parity-even class of models was analyzed with $\overline{a}_I = 0$, $\overline{b}_J = 0$. The available results are summarized in Fig.~1.

\subsection{Torsion-free vacuum solutions}

Let us consider the vacuum solutions with {\it vanishing torsion} in the general quadratic models (\ref{LRT}). In vacuum, the matter sources vanish, ${\mathfrak T}_\alpha = 0$ and ${\mathfrak S}_{\alpha\beta} = 0$, and for $T^\alpha=0$ we find, after some straightforward algebra, that the curvature scalar is constant,
\begin{equation}
\widetilde{R} = -\,{\frac {4\lambda_0} {a_0}},\label{scalR}
\end{equation}
whereas the general field equations (\ref{ERT1}) and (\ref{ERT2}) reduce to 
\begin{eqnarray}
(b_1 + b_4)\,{}^{*(1)}\!\widetilde{R}_{\alpha\beta}\wedge\widetilde{\Psi}^\beta
+ (\overline{b}_1 - \overline{b}_4)\,{}^{(1)}\!\widetilde{R}_{\alpha\beta}\wedge
\widetilde{\Psi}^\beta &=& \widehat{a}_0\,{}^*\widetilde{\Psi}_\alpha, \label{firstR}\\
\left[(b_1 + b_4)^2 + (\overline{b}_1 - \overline{b}_4)^2\right]\,\widetilde{D}
\,\widetilde{\Psi}_\alpha &=& 0.\label{secondR}
\end{eqnarray}
The tilde denotes the torsion-free Riemannian objects and operators. Here we defined 
\begin{equation}
\widehat{a}_0 = a_0 - {\frac {2\lambda_0}{3a_0}}\,(b_4 + b_6).\label{alpha}
\end{equation}
The 1-form $\Psi_\alpha$ introduced in (\ref{Psia}), determines the structure of the fourth irreducible part of the curvature ${}^{(4)}\!R_{\alpha\beta} = - \vartheta_{[\alpha}\wedge\Psi_{\beta]}$; its components coincide with the symmetric traceless Ricci tensor (\ref{RicP}). 

Clearly, all Einstein spaces, i.e., the solutions of the vacuum Einstein equations with a cosmological term
\begin{equation}
\widetilde{\Psi}_\alpha = 0,\label{ein}
\end{equation}
recall (\ref{RicP}), are vacuum torsion-free solutions of (\ref{firstR})-(\ref{secondR}) in the general quadratic Poincar\'e gauge models. Actually, a stronger result can be demonstrated.

{\sl Theorem}. The Einstein spaces (\ref{ein}) are {\it the only} torsion-free vacuum solutions of (\ref{firstR})-(\ref{secondR}) for {\it all values  of the coupling constants} except for the three very specific degenerate choices:
\begin{equation}
b_6 - {\frac {3a_0^2}{2\lambda_0}} = \left\{\begin{array}{c} 
b_1, \\ - b_4,\\ -2b_1 - 3b_4.\end{array}\right.\label{except}  
\end{equation}
To begin the proof, we notice that if $b_1 + b_4 = 0$ and $\overline{b}_1 - \overline{b}_4 = 0$, the system (\ref{firstR}) and (\ref{secondR}) reduces to (\ref{ein}). Now, we assume that  $b_1 + b_4 \neq 0$ and $\overline{b}_1 - \overline{b}_4 \neq 0$. Taking the covariant exterior derivative of (\ref{secondR}), we then find $\widetilde{D}\widetilde{D}\,\widetilde{\Psi}_\alpha = -\,\widetilde{R}_\alpha{}^\beta\wedge\widetilde{\Psi}_\beta = -\,{}^{(1)}\!\widetilde{R}_\alpha{}^\beta\wedge\widetilde{\Psi}_\beta = 0$. Consequently, the second term on the left-hand side of (\ref{firstR}) disappears and the system (\ref{firstR}) and (\ref{secondR}) is recast into
\begin{eqnarray}
{}^{*(1)}\!\widetilde{R}_{\alpha\beta}\wedge\widetilde{\Psi}^\beta
= \Lambda\,{}^*\widetilde{\Psi}_\alpha, \qquad 
\widetilde{D}\,\widetilde{\Psi}_\alpha = 0,\label{vacuumTF}
\end{eqnarray}
where we put 
\begin{equation}
\Lambda = {\frac {\widehat{a}_0}{b_1 + b_4}}.\label{vacL}
\end{equation} 
The final step is technically nontrivial, and the value of $\Lambda$ is crucial. A direct analysis making use of the Newman-Penrose technique \cite{OPZ} (see also the earlier works \cite{Debney,Fair1,Fair2}) shows that (\ref{ein}) is the only solution of the system (\ref{vacuumTF}) for all values of $\Lambda$, except for the three cases when $3\Lambda/2 = \{0, \widetilde{R}/4, -\widetilde{R}/2\}$. Using then (\ref{alpha}), (\ref{vacL}) and (\ref{scalR}), we prove (\ref{except}).

\subsection{Gravitational planes waves in Poincar\'e gravity}

Gravitational waves are of fundamental importance in physics, and recently the purely theoretical research in this area was finally supported by the first experimental evidence. The plane-fronted gravitational waves represent an important class of exact solutions which generalize the basic properties of electromagnetic waves in flat spacetime to the case of curved spacetime geometry.

To streamline the presentation, we put the cosmological constant $\lambda_0 = 0$ here.

\subsubsection{Electromagnetic plane waves}

The key for the description of a plane wave on a spacetime manifold is the null shear-free geodetic covector field. More exactly, one talks of the wave 1-form $k = d\varphi$ which arises from the phase function $\varphi$ (so that the wave covector is $k_\alpha = e_\alpha\rfloor k$) with
the properties
\begin{eqnarray}
k\wedge{}^\ast\!k = 0,\qquad k\wedge{}^\ast\!Dk^\alpha = 0,\label{kk1}\\
k\wedge{}^\ast\!F = 0,\qquad k\wedge F = 0,\qquad F\wedge{}^\ast\!F = 0.\label{kF1}
\end{eqnarray}

Here $F$ is the electromagnetic field strength 2-form. The actual structure of the wave configurations depends on the Lagrangian of the electromagnetic field. For example, in Maxwell's theory in the flat Minkowski spacetime (specializing to the case $e^\alpha_i = \delta^\alpha_i, \Gamma_i{}^{\alpha\beta} = 0$) the electromagnetic plane wave is given by
\begin{equation}
F = k\wedge a,\qquad k\wedge{}^\ast\!a = 0.\label{Fij}
\end{equation}
Here the wave covector is constant, $dk = 0$, whereas the polarization 1-form $a$ depends only 
on the phase, $a_i = a_i(\varphi)$ and satisfies the above orthogonality relation.

\subsubsection{Gravitational plane waves}

In order to discuss the gravitational wave solutions in Poincar\'e gravity theory, we start with the flat Minkowski geometry described by the coframe and connection $\widehat{\vartheta}^\alpha = dx^\alpha$, $\widehat{\Gamma}^{\alpha\beta} = 0$. Introducing the phase variable $\sigma = x^0 - x^1$, we construct the wave 1-form $k = d\sigma = \widehat{\vartheta}^{\hat 0} - \widehat{\vartheta}^{\hat 1}$. The gravitational wave ansatz then reads
\begin{eqnarray}
\vartheta^\alpha &=& \widehat{\vartheta}^\alpha + {\frac 12}U\,k^\alpha k,\label{cofW}\\ 
\Gamma^{\alpha\beta} &=& \widehat{\Gamma}^{\alpha\beta} 
+ (k^\alpha W^\beta - k^\beta W^\alpha)k.\label{gamW}
\end{eqnarray}
Importantly, this ansatz does not change the wave 1-form which is still defined by
\begin{equation}
k = d\sigma = \vartheta^{\widehat 0} - \vartheta^{\widehat 1}.\label{kdef}
\end{equation}
By construction, we have $k\wedge{}^\ast\!k = 0$. The wave covector is constructed as usual as $k_\alpha = e_\alpha\rfloor k$, so that its (anholonomic) components are $k_\alpha = (1, -1, 0, 0)$ and $k^\alpha = (1, 1, 0, 0)$. Hence, this is a null vector field, $k_\alpha k^\alpha = 0$. 

The two unknown variables $U$ and $W_\alpha$ determine the wave profile, and we choose them as functions $U = U(\sigma, x^A)$ and $W^\alpha = W^\alpha(\sigma, x^A)$. Here $x^A = (x^2,x^3)$, from now on the indices from the beginning of the Latin alphabet $a,b,c... = 0,1$, whereas the capital Latin indices run $A,B,C... = 2,3$. In addition, we assume the orthogonality $k_\alpha W^\alpha = 0$, which is guaranteed if we choose 
\begin{equation}\label{Wa0}
W^\alpha = \begin{cases}W^a = 0,\qquad\qquad\qquad a = 0,1, \\
W^A = W^A(\sigma, x^B),\qquad A = 2,3.\end{cases}
\end{equation}

The resulting line element then reads (with $\rho = x^0 + x^1$)
\begin{equation}
ds^2 = d\sigma d\rho + Ud\sigma^2 - \delta_{AB}dx^Adx^B.\label{ds_2}
\end{equation}

In view of the properties of the objects defined above, we verify that the wave 1-form is closed, and the wave covector is constant:
\begin{equation}
dk = 0,\qquad dk_\alpha = 0,\qquad Dk_\alpha = 0.\label{dk0}
\end{equation}
Taking this into account, we straightforwardly compute the torsion and the curvature 2-forms:
\begin{eqnarray}
T^\alpha = k\wedge a^\alpha,\qquad R^{\alpha\beta} = k\wedge a^{\alpha\beta},\label{curW}
\end{eqnarray}
where we introduced the 1-forms 
\begin{eqnarray}
a^\alpha &=& -\, k^\alpha\Theta,\qquad \Theta := {\frac 12}\,\underline{d}\,U 
+ W_\alpha\vartheta^\alpha,\label{TMW}\\
a^{\alpha\beta} &=& -\,2k^{[\alpha}\Omega^{\beta]},\qquad \Omega^\alpha 
:= \underline{d}\,W^\alpha. \label{OMW}
\end{eqnarray}
The differential $\underline{d}$ acts in the transversal 2-space spanned by $x^A = (x^2, x^3)$.

It is worthwhile to notice that the 2-forms of the gravitational Ponicar\'e gauge field strengths (\ref{curW}) have the same structure as the electromagnetic field strength (\ref{Fij}) of a plane wave. Now $a^\alpha$ and $a^{\alpha\beta}$ play the role of the gravitational (translational and rotational, respectively) ``polarization'' 1-forms. In complete analogy to the polarization 1-form $a$ in (\ref{Fij}), we notice that the gravitational polarization 1-forms satisfy the orthogonality relations
\begin{eqnarray}
k\wedge{}^\ast a^\alpha = 0,\qquad k\wedge{}^\ast a^{\alpha\beta} = 0. \label{aka1}
\end{eqnarray}
Clearly, the gravitational field strengths of a wave have the properties
\begin{eqnarray}\label{kTW}
k\wedge{}^\ast\!T^\alpha = 0,\qquad &k\wedge T^\alpha = 0,&\qquad T^\alpha\wedge{}^\ast\!T^\beta = 0,\\
k\wedge{}^\ast\!R^{\alpha\beta} = 0,\qquad &k\wedge R^{\alpha\beta} = 0,&
\qquad R^{\alpha\beta}\wedge{}^\ast\!R^{\rho\sigma} = 0,\label{kRW}
\end{eqnarray}
in complete analogy to the electromagnetic plane wave (\ref{kF1}).

In addition, however, the gravitational Ponicar\'e gauge field strengths satisfy 
\begin{equation}
k_\alpha\,T^\alpha = 0,\qquad k_\alpha\,R^{\alpha\beta} = 0.\label{kTRW2}
\end{equation}

The explicit gravitational wave solution is constructed as follows. The wave profile vector variable is expressed in terms of potentials
\begin{equation}\label{pW}
W^A = {\frac 12}\delta^{AB}\partial_B(U + V) + {\frac 12}\eta^{AB}\partial_B\overline{V},
\end{equation}
where $\eta^{AB} = -\,\eta^{BA}$ is the totally antisymmetric Levi-Civita tensor on the 2-dimensional space of the wave front. Substituting the wave ansatz (\ref{cofW}), (\ref{gamW}) and (\ref{pW}) into (\ref{ERT1}) and (\ref{ERT2}), the highly nonlinear system of the gravitational field equations quite remarkably reduces to the system of three linear differential equations
\begin{equation}
\underline{\Delta}\,{\cal V} - M\,{\cal V} = 0,\label{DSVc}
\end{equation}
for the wave profile potentials which are conveniently assembled in a column ``3-vector'' variable  ${\cal V} = \left(\begin{array}{c} U\\ V \\ \overline{V}\end{array}\right)$. Here $\underline{\Delta} = \delta^{AB}\partial_A\partial_B$ is the 2-dimensional Laplacian on the $(x^2,x^3)$ space, and the $3\times 3$ matrix $M$ is constructed from the coupling constants $a_I, \overline{a}_I, b_J, \overline{b}_J$. One can straightforwardly solve the system (\ref{DSVc}) by diagonalizing $M$. 

Remarkably, eigenvalues of $M$ coincide \cite{PGW1,PGW2} with the masses of the particle spectrum of the propagating torsion modes in quadratic PG models \cite{Hay,Nev1,Nev2,Sez1,Sez2,Kuh,Karananas}. The results above can be further generalized to $\lambda_0\neq 0$ by using a modified ansatz (\ref{cofW})-(\ref{gamW}) with ($\widehat{\vartheta}^\alpha, \widehat{\Gamma}{}^{\alpha\beta}$) describing the de Sitter geometry \cite{PGW2}.

\section{Cosmology in Poincar\'e gravity}

Taking into account the spin of matter (as a new physical source of gravity) and the torsion (as an additional geometrical property of spacetime), leads to modifications of early and late stages of universe's evolution \cite{Trautman:1973,Mink1,Magu,Pop,Dirk}. This potentially contributes to the solution of the two important issues of the modern cosmology: the singularity problem and the dark energy problem.

The cosmological evolution in the Einstein-Cartan theory (\ref{LHE}) and in the most general quadratic models (\ref{LRT}) are qualitatively different. This is due to the fact that in the former case the torsion is not dynamical and can be eliminated. 

Let us consider the Friedman-Robertson-Walker (FRW) geometry with the spacetime interval
\begin{equation}
ds^2 = (dx^0)^2 - {\frac {a^2}{\left(1 + {\frac {r^2}{4\ell^2}}\right)^2}}
\left\{(dx^1)^2 + (dx^2)^2 + (dx^3)^2\right\}.\label{dsFRW}
\end{equation}
The local coordinates are $x^i = \{x^0, x^1, x^2, x^3\}$, and $r^2 = (x^1)^2 + (x^2)^2 + (x^3)^2$. Here the scale factor depends on the cosmological time, $a = a(x^0)$, and the parameter $\ell^2$ determines the geometry of the 3-space. The latter is the space of constant curvature $k = {\frac 1{\ell^2}}$ which can be zero ($k = 0$: flat space), positive ($k > 0$: closed space) or negative ($k < 0$: open space). The geometry of this 3-space is described by the coframe and the Riemannian local Lorentz connection
\begin{equation}
\underline{\vartheta}^a = {\frac {dx^a}{1 + {\frac {r^2}{4\ell^2}}}},\qquad 
\underline{\Gamma}{}^{ab} = {\frac {1}{2\ell^2}}\left(x^a\underline{\vartheta}^b
- x^b\underline{\vartheta}^a\right).\label{CG3}
\end{equation}
The Latin indices from the beginning of the alphabet run $a,b,c,\dots = 1,2,3$ and they are raised and lowered with the help of the Euclidean metric $\delta_{ab}$ and $\delta^{ab}$. For example, $x_b = \delta_{ab}x^a$ and $\underline{\vartheta}_b = \delta_{ab} \underline{\vartheta}^a$.

\subsection{Einstein-Cartan cosmology}

In cosmology, it is common to use the hydrodynamic description of matter. An appropriate model is Weyssenhoff spinning fluid with the canonical energy-momentum and spin currents (\ref{TW}) and (\ref{SW}). Since the second field equation (\ref{EC2}) is algebraic, we can use it to express the torsion as a linear function of spin. Substituting the torsion into the first field equation (\ref{EC1}), we then recast it into the Einstein equation ${\frac 12}\eta_{\alpha\beta\gamma}\wedge \widetilde{R}^{\beta\gamma} = \kappa\,{\mathfrak T}^{\rm eff}_\alpha$ with the effective energy-momentum current
\begin{eqnarray}
{\mathfrak T}^{\rm eff}_\alpha = -\,p^{\rm eff}\big(\eta_\alpha - {\frac 1{c^2}}u_\alpha u\big) 
+ {\frac {\varepsilon^{\rm eff}}{c^2}}\,\eta_\alpha + 
\Big(g^{\nu\lambda} + {\frac 1{c^2}}u^\nu u^\lambda\Big)\widetilde{D}_\nu\left(
u_{(\mu}{\mathcal S}_{\alpha)\lambda}\right)\eta^\mu,\label{QTeWF}
\end{eqnarray}
where the effective pressure and energy density depend on spin:
\begin{eqnarray}
p^{\rm eff} = p -  {\frac {\kappa c^2{\mathcal S}^2}{4}},\qquad
\varepsilon^{\rm eff} = \varepsilon -  {\frac {\kappa c^2{\mathcal S}^2}{4}}.\label{PEeff}
\end{eqnarray}

In order to have a qualitative understanding of the Einstein-Cartan cosmology, let us specialize to the case of the flat model (with $k = 0$) for the dust equation of state $p = 0$. For the FRW ansatz (\ref{dsFRW}), the effective Einstein equation then reduces to the generalized Friedman equation
\begin{equation} 
3\,{\frac {\dot{a}^2}{a^2}} = \varepsilon^{\rm eff},\qquad \varepsilon^{\rm eff}= \varepsilon 
- {\frac {\kappa c^2{\mathcal S}^2}{4}}.\label{ECFried}
\end{equation}
The conservation laws of the energy-momentum (\ref{conmomC}) and spin (\ref{dS0}) yield
\begin{equation}\label{ESdust}
\varepsilon = {\frac {\varepsilon_0}{a^3}},\qquad {\mathcal S} = {\frac {{\mathcal S}_0}{a^3}}.
\end{equation}
The equation (\ref{ECFried}) can be straightforwardly integrated, and we observe that the cosmological evolution is nonsingular \cite{Trautman:1973}. At the time of a bounce, the universe occupies a minimal volume, when the energy density is maximal:
\begin{equation}
a_{\rm min}^3 = {\frac {\kappa c^2{\mathcal S}_0^2}{4\varepsilon_0}},\qquad
\varepsilon_{\rm max} = {\frac {\varepsilon_0}{a_{\rm min}^3}} = 
{\frac {4\varepsilon_0^2}{\kappa c^2{\mathcal S}_0^2}}.\label{amin} 
\end{equation}
One can evaluate the latter by assuming that the cosmological dust matter is composed of fermions with a mass $m$ and spin $\hbar/2$. Then the ratio of the energy density per spin density is $\varepsilon_0/{\mathcal S}_0 = 2mc^2/\hbar$. As a result, we find 
\begin{equation}
\varepsilon_{\rm max} = {\frac {16m^2c^2}{\kappa \hbar^2}}.\label{Emax1}
\end{equation}
For the mass of a nucleon, the corresponding maximal {\it mass density} is thus ${\frac {\varepsilon_{\rm max}}{c^2}} = {\frac {2m^2c^4}{\pi G\hbar^2}} \approx 10^{57}$~kg/m$^3$. At the late stage, when the first term  $\sim 1/a^3$ on the right-hand side of the generalized Friedman equation (\ref{ECFried}) becomes dominating over the second $\sim 1/a^6$ term, the evolution of the scale factor approaches the usual law $a(x^0)\sim (x^0)^{\frac 23}$ of the dust FRW cosmology.

\subsection{Cosmology in general Poincar\'e gauge gravity}

In contrast to the Einstein-Cartan theory, in the general quadratic Poincar\'e gauge gravity models (\ref{LRT}) the torsion degrees of freedom are propagating. Accordingly, one has to come up with an appropriate description of the torsion. 

We construct the generalized FRW cosmology (\ref{dsFRW}) in the  Poincar\'e gauge gravity theory with the help of the ansatz for the coframe $\vartheta^\alpha$ and connection $\Gamma^{\alpha\beta}$:
\begin{eqnarray}
\vartheta^{\hat{0}} = dx^0,&\qquad& \vartheta^a = a\,\underline{\vartheta}^a,\label{cofFRW}\\
\Gamma^{{\hat{0}}a} = b\,\underline{\vartheta}^a,&\qquad& \Gamma^{ab} = \underline{\Gamma}{}^{ab}
- \sigma\,\epsilon^{ab}{}_c\underline{\vartheta}^c.\label{gamFRW}
\end{eqnarray}
This configuration is described by the three functions of the cosmological time
\begin{equation}
a = a(x^0),\qquad b = b(x^0),\qquad \sigma = \sigma(x^0).\label{abs}
\end{equation}
For the torsion we then find ${}^{(1)}\!T^\alpha = 0$, whereas
\begin{equation}
{}^{(2)}\!T^a = v\,\vartheta^{\hat{0}}\wedge\vartheta^a,\qquad
{}^{(3)}\!T^a = \overline{v}\,\epsilon^a{}_{bc}\,\vartheta^b\wedge\vartheta^c,\label{iTcos}
\end{equation}
where we denoted
\begin{equation}
v = {\frac {\dot{a} - b}{a}},\qquad \overline{v} = {\frac {\sigma}{a}}.\label{vv}
\end{equation}

One can show that the field equations (\ref{ERT1})-(\ref{ERT2}) allow only for a spinless matter with ${\mathcal S}_{\alpha\beta} = 0$, and thus the Weyssenhoff medium reduces to the ideal fluid. In order to compare the resulting dynamics to the Einstein-Cartan cosmology, we specialize to the case when the cosmological constant vanishes $\lambda_0 = 0$, the spatial geometry is flat $k = 0$, and the cosmological matter has the equation of state of a dust $p = 0$. To simplify computations, we also assume $a_2 = 0$ and limit ourselves to the class of parity-even models with $\overline{a}_I = 0$, $\overline{b}_J = 0$. Then we find that the axial torsion vanishes $\overline{v} = 0$, whereas $v$ turns out to be proportional to the Hubble function $\dot{a}/a$, and the system (\ref{ERT1})-(\ref{ERT2}) reduces to the generalized Friedman equation
\begin{equation}
3a_0{\frac {\dot{a}^2}{a^2}} = \kappa\varepsilon_{\rm eff},\qquad
\varepsilon_{\rm eff} = \varepsilon\,{\frac {\left(1 - {\frac {\varepsilon}{\varepsilon_\ell}}
\right)\left(1 - {\frac {\varepsilon}{2\varepsilon_\ell}}\right)}{\left(1 + 
{\frac {\varepsilon}{\varepsilon_\ell}}\right)^2}},\qquad 
\varepsilon_\ell := {\frac {12a_0^2}{\kappa\ell_\rho^2(b_4 + b_6)}}.\label{epsell}
\end{equation}
Taking into account the explicit dependence of the energy density on the scale factor (\ref{ESdust}), we can integrate the  generalized Friedman equation, and the solution for $a = a(x^0)$ is expressed in terms of the elliptic integrals. The qualitative result is as follows. The cosmological evolution is again non-singular \cite{Mink1,Mink2}, with the minimal value of the scale factor and the highest energy density:
\begin{equation}\label{aminQR}
 a_{\rm min}^3 = {\frac {\kappa\varepsilon_0\ell_\rho^2(b_4 + b_6)}{12a_0^2}},\qquad
\varepsilon_{\rm max} = {\frac {\varepsilon_0}{a_{\rm min}^3}} 
= {\frac {12a_0^2}{\kappa\ell_\rho^2(b_4 + b_6)}}.
\end{equation}
At the late stages of the cosmological evolution, when the universe expands to sufficiently large values of the scale factor so that the condition ${\frac {\varepsilon}{\varepsilon_\ell}} = {\frac {\varepsilon_0}{\varepsilon_\ell a^3}} \ll 1$ is satisfied, eq. (\ref{epsell}) reduces to the usual Friedman equation for the dust matter, and the evolution law asymptotically is approximated by the law  $a(x^0)\sim (x^0)^{\frac 23}$. 

Let us make a blitz comparison of the Einstein-Cartan model and the general quadratic PG model. Both models predict a non-singular cosmological scenario which at the later stage approaches the standard Friedman evolution. However, the values of $a_{\rm min}$ and $\varepsilon_{\rm max}$ are determined differently: whereas in the Einstein-Cartan theory the parameters of the bounce (\ref{amin}) depend on the spin of matter, in the general quadratic model the properties of matter are irrelevant and the values (\ref{aminQR}) are determined by the universal strong gravity coupling constant $1/\rho$ and the corresponding new length scale $\ell_\rho^2$. 

Including odd-parity terms in the Lagrangian (with the nontrivial constants $\overline{a}_I$ and $\overline{b}_J$), and allowing for the odd-parity torsion $\overline{v}$, the cosmological equations are extended to a highly nontrivial system for the the scale factor $a$, and the torsion functions $v$ and $\overline{v}$. In general, the space of solutions for this system encompasses both the non-singular and singular cosmological scenarios, see \cite{Chen,Ho2,Ho3}, e.g.

\section{Motion of test bodies in Poincar\'e gauge gravity}

Before discussing the dynamics of massive extended bodies in the gravitational field, it is useful to recall the electromagnetism. An electrically charged body is characterized by the electric current density $J^\alpha$ which describes how the charges and currents are distributed inside this body. When the size of the body is much smaller than the typical length over which the electric and magnetic fields change significantly, it can be treated as a test particle. Choosing a reference point $y^\alpha$ inside the body, one interprets the curve $y^\alpha = y^\alpha(\tau)$ as the world line of the body with the velocity $u^\alpha = dy^\alpha/d\tau$, and introduces a set of the multipole moments as integrals $\int_\Sigma \delta x^{\mu_1}\cdots \delta x^{\mu_n} J^\alpha$ over a spatial cross-section $\Sigma$ of the world tube swept by the body through its motion in the spacetime, where $\delta x^\mu = x^\mu - y^\mu$ gives the position of charged material elements relative to the reference point. The lowest moments are the total electric charge of a body, its electric dipole moment and so on. 

Qualitatively, in this approach an extended body is replaced by a test particle characterized by (infinite number of) multipoles which describe the internal structure of the body and contain all the information which was encoded in the electric current $J^\alpha$. In a similar way, in order to analyse the motion of a massive body in the gravitational field, one needs to take the corresponding gravitational matter currents and to construct the multipole moments for them. This technique has a long history going back to Einstein, Weyl, Infeld, Mathisson, Papapetrou, Dixon (for the historic introduction and key references, see \cite{eom1,eom2}, for example). Remarkably, the equations of motion of the multipole moments should not be postulated, but they follow directly from the conservation laws of the Noether currents. 

In Einstein's GR, the gravitational field couples to the energy-momentum current of the structureless matter. This corresponds to the group of spacetime translations (diffeomorphisms) which underlies GR. The Poincar\'e gauge gravity takes into account a possible nontrivial microstructure of matter and extends the theory to the Poincar\'e current $({\mathfrak T}_\alpha, {\mathfrak S}_{\alpha\beta})$ which includes the translational {\it and} Lorentz currents, i.e., the canonical energy-momentum and the spin of matter. Accordingly, we will have two types of multipoles.

There exist many multipole expansion schemes (both noncovariant and covariant) in gravity theory. Among them, the most convenient one is the covariant expansion technique based on Synge's world-function formalism \cite{Synge}, first used by Dixon \cite{Dixon} to define a set of moments characterizing the test body. The world-function $\sigma(x,y)$ measures the length of the geodesic curve connecting the spacetime points $x$ and $y$. Using a condensed notation when tensor indices are labeled by spacetime points to which they are attached, we denote by $\sigma_y:= \widetilde{\nabla}_y \sigma$ a covariant derivative of the world-function. The parallel propagator by $g^y{}_x(x,y)$ describes the parallel transport of objects along the unique geodesic that links the points $x$ and $y$, e.g.: given a vector $V^x$ at $x$, the corresponding vector at $y$ is obtained by means of the parallel transport along the geodesic curve as $V^y = g^y{}_x(x,y)V^x$. For more details, see \cite{Synge,eom2}. In PG theory, we define the multipole moments of arbitrary order:
\begin{eqnarray}
cp_{y_1\dots y_n y_0} \!\!&:=&\!\! (-1)^n\!\!\int\limits_{\Sigma(\tau)}\!\!\sigma_{y_1}\cdots
\sigma_{y_n}g_{y_0}{}^{x_0}{\mathfrak T}_{x_0}{}^{x_1}d\Sigma_{x_1},\label{pmom}\\
s_{y_2\dots y_{n+1}y_0 y_1} \!\!&:=&\!\! (-1)^n\!\!\int\limits_{\Sigma(\tau)}\!\!\sigma_{y_2}\cdots
\sigma_{y_{n+1}}g_{y_0}{}^{x_0} g_{y_1}{}^{x_1}{\mathfrak S}_{x_0 x_1}{}^{x_2}d\Sigma_{x_2},\label{hmom}\\
q_{y_3\dots y_{n+2}y_0 y_1}{}^{y_2} \!\!&:=&\!\! (-1)^n\!\!\int\limits_{\Sigma(\tau)}\!\!\sigma_{y_3}\cdots
\sigma_{y_{n+2}} g_{y_0}{}^{x_0} g_{y_1}{}^{x_1} g^{y_2}{}_{x_2}{\mathfrak S}_{x_0 x_1}{}^{x_2}w^{x_3}
d\Sigma_{x_3}.\label{qmom}
\end{eqnarray}

With these definitions, the equations of motion of test bodies are derived by integrating the conservation laws of the energy-momentum (\ref{conmomC}) and angular momentum (\ref{Noe2}) over the cross-section of the world tube. The resulting system describes the dynamics of multipole moments (\ref{pmom}) and (\ref{hmom}) of any order. In the {\it pole-dipole approximation}, we find
\begin{eqnarray}
{\frac {\widetilde{D}{\cal P}_\alpha}{d\tau}} &=& {\frac 12}\widetilde{R}_{\alpha\beta}{}^{\mu\nu}
u^\beta{\cal J}_{\mu\nu} - {\frac 12}Q^{\mu\nu}{}_{\beta}\widetilde{\nabla}_\alpha T_{\mu\nu}{}^{\beta} ,\label{DPdec}\\
{\frac {\widetilde{D}{\cal J}_{\alpha\beta}}{d\tau}} &=& {\cal P}_\alpha u_\beta - {\cal P}_\beta u_\alpha 
- Q_{\mu\nu[\alpha}T^{\mu\nu}{}_{\beta]} - 2Q_{[\alpha|\mu\nu|}T_{\beta ]}{}^{\mu\nu}.\label{DJdec}
\end{eqnarray}
Here ${\frac {\widetilde{D}}{d\tau}} = u^\alpha\widetilde{\nabla}_\alpha$ is the Riemannian covariant derivative with respect to the proper time $\tau$, $u^\alpha$ is body's 4-velocity. We defined the generalized total energy-momentum 4-vector and the generalized total angular momentum (with the contortion tensor $K_{i}{}^{\alpha\beta} := \widetilde{\Gamma}_{i}{}^{\alpha\beta} - \Gamma_{i}{}^{\alpha\beta}$) by
\begin{eqnarray}
{\cal P}_\alpha := p_\alpha - {\frac 12}K_\alpha{}^{\mu\nu}s_{\mu\nu},\qquad
{\cal J}_{\alpha\beta} := p_{\alpha\beta} - p_{\beta\alpha} + s_{\alpha\beta}.\label{PJtot}
\end{eqnarray}
and introduced $Q_{\alpha\beta\mu} := {\frac 12}\left(q_{\alpha\beta\mu} + q_{\alpha\mu\beta} 
- q_{\beta\mu\alpha}\right)$.

In the {\it monopole approximation}, the equations of motion are simplified to
\begin{equation}
{\frac {\widetilde{D}{\cal P}_\alpha}{d\tau}} = 0,\qquad {\cal P}_\alpha u_\beta - {\cal P}_\beta u_\alpha
= 0.\label{mono}
\end{equation}
Hence we find ${\cal P}_\alpha = Mu_\alpha$, where $Mc^2 = {\cal P}_\alpha u^\alpha$, and therefore
(\ref{mono}) reduces to the Riemannian geodesic ${\frac {\widetilde{D}u^\alpha}{d\tau}} = 
u^\beta\widetilde{\nabla}_\beta u^\alpha = 0$.

\section{Conclusion: Probing spacetime geometry}

Einstein \cite{Ein1921} underlined: ``...The question whether this continuum has a Euclidean, Riemannian, or any other structure is a question of physics proper which must be answered by experience, and not a question of a convention to be chosen on grounds of mere expediency." How can one probe a possible deviation of the spacetime structure from the Riemannian geometry?

One needs matter (particles, bodies, continua, fields) with microstructure, i.e., with intrinsic spin \cite{Yasskin,eom1}. This is most clearly seen from the equations of motion above. Contrary to some unfortunate statements in the literature: (i) only the spin --not the orbital rotation-- couples to the non-Riemannian geometry and hence only from observations of the spin dynamics one can measure (or set bounds on) the spacetime torsion, (ii) a massive test point (monopolar body) {\it always} moves along the Riemannian geodesic and not along the non-Riemannian autoparallel. 

So far, there is no evidence of the torsion in nature. Following the early theoretical work \cite{Adam,Rumpf,Hay1} on the precession of spin in the Riemann-Cartan spacetime, the upper limit $|\,T\,| < 10^{-15} {\frac 1 {\rm m}}$ was established for the torsion from Huges-Drever type experiments \cite{Lamm}, from the analysis of the Lorentz violations in Standard Model extensions \cite{Kost}, from the search of spin-spin interaction using Earth as a test body \cite{Hunter}, as well as from the study of the nuclear spin dynamics \cite{ostor}; milder constraints $|\,T\,| < 10^{-2} {\frac 1 {\rm m}}$ were derived from the precession of neutron's spin in liquid $^4$He.

\section*{Acknowledgments}

This work was partially supported by the Russian Foundation for Basic Research (Grant No. 16-02-00844-A). I am grateful to Friedrich Hehl and Dirk Puetzfeld for the most useful comments and advice.

\end{document}